\documentclass[10pt]{iopart}

\usepackage{graphicx}
\usepackage[usenames]{color}
\usepackage{amssymb}

\begin{document}


\title{Vortex lattice  in a uniform Bose-Einstein condensate in a box trap}

\author{ S. K. Adhikari}
 
\address{
Instituto de F\'{\i}sica Te\'orica, UNESP - Universidade Estadual Paulista, 01.140-070 S\~ao Paulo, S\~ao Paulo, Brazil
} 
\ead{sk.adhikari@unesp.br}



%

\begin{abstract}

We study numerically the vortex-lattice formation in a rapidly rotating uniform quasi-two-dimensional Bose-Einstein condensate (BEC) in a box trap. We consider two types of boxes: square and circle.  In a square-shaped 2D box trap, when the number of generated vortices is the square of an integer, the vortices are found to be arranged in a perfect square lattice, although deviations near the center are found when the number of generated vortices is arbitrary. In case of a  circular box trap, the generated vortices in the rapidly rotating BEC 
lie  on concentric closed orbits. Near the center, these orbits have the shape of polygons, whereas near the  
periphery the orbits are circles. The circular box trap is equivalent to the rotating 
cylindrical bucket used in early experiment(s)  with liquid He II.
The number of generated vortices in both cases   is in qualitative agreement with Feynman's universal estimate. 
 The numerical simulation for this study is performed by a solution of the underlying mean-field Gross-Pitaevskii (GP) equation in the rotating frame, where the wave function for the generated vortex lattice is a stationary state. Consequently, the imaginary-time propagation method can be used for a solution of the GP equation, known to lead to an accurate numerical solution. 
We also demonstrated the dynamical stability of the vortex lattices in real-time propagation upon a small change of the angular frequency of rotation, using the converged imaginary-time wave function as the initial state.

\end{abstract}




\newpage

\section{Introduction}

 Soon after the observation  of a trapped Bose-Einstein condensate (BEC) in alkali-metal atoms  at ultra-low density and temperature in a laboratory \cite{becexpt,becexpt2}, rotating 
trapped condensates  were created in laboratory under controlled conditions  and studied experimentally \cite{vors}. 
A small number of vortices were created \cite{vors} for small values of the angular frequency of rotation $\Omega$. With the increase of  $\Omega$,  vortex arrays containing a large number of vortices 
were generated \cite{vorl}. As suggested by Onsager \cite{onsager}, Feynman \cite{feynman} and Abrikosov \cite{abri}
 these vortices have quantized circulation as in liquid He II: \cite{Sonin,fetter}
\begin{equation}\label{qv}
\oint_{\cal C} {\bf v} .  d{\bf r}=\frac{2\pi\hbar l}{m},
\end{equation}
where ${\cal C}$ is a generic closed path, ${\bf v}({\bf r},t)$ is the super-fluid velocity field,  
  $l$ is  the quantized integral  angular momentum of an atom in units of $\hbar$ in the rotating BEC and $m$ is the mass of an atom. The generation of quantized vortices in a dilute BEC or in liquid He II upon rotation is an earmark of super-fluidity. Nevertheless,  the condensate fraction in He II is very small ($\sim 10\%$) \cite{leggett} compared to the condensate fraction in a weak-coupling dilute BEC (close to $ 100\%$)
 \cite{smith}. This makes the generation of a clean vortex lattice in rotating He II
next to impossible, whereas,  in a rotating dilute trapped BEC, clean vortex lattices of more than 100 vortices have been observed \cite{vorl}.  
As the angular frequency of rotation is increased in the rotating BEC,   energy consideration favors the  formation of  a lattice of quantum vortices of unit angular momentum  each ($l=1$) \cite{abri,fetter} and not angular momentum states with $l>1$. This was first confirmed experimentally  in liquid He II in bulk \cite{4he} and later in dilute trapped BEC \cite{vors,vorl}.
Consequently, a rapidly rotating trapped BEC is found to form a large number of vortices of unit angular momentum  arranged with a definite symmetry usually  in a Abrikosov  triangular lattice \cite{vorl,abri}.     
In the weak-coupling low-density limit of the trapped rotating BEC, it has been possible to study the formation of  vortices  by the mean-field Gross-Pitaevskii (GP) equation \cite{gross} in the rotating frame   using imaginary-time \cite{theoryIT} and real-time \cite{theoryRT} propagation.

More recently it has been possible to create a BEC in a laboratory in a one- \cite{1dbox} and a multi-dimensional \cite{3dbox} optical  box trap so that the condensate is subjected to a uniform potential inside the box. The created  trap in Ref. \cite{1dbox} provides a  uniform confinement in one direction
using lasers and  strong magnetic   confinements in the transverse directions. A two-dimensional (2D)
square  box trap can be prepared with two such orthogonal optical  1D traps   in the $x-y$ plane with a strong trap in the transverse $z$ direction. 
The optical  trap considered in Ref. \cite{3dbox}    is a uniform three-dimensional  cylindrical  box trap  which  combines   a circular box trap in the $x-y$ plane and a uniform 1D confinement in the $z$ direction.    
The experimental technique thus allows the possibility of the creation of a uniform BEC in a 2D quantum box 
potential using laser beams and a strong magnetic confinement in the transverse direction, so that an effectively uniform quasi-two-dimensional
 (quasi-2D) BEC is generated  in the form of a square or a circle.  
To create a uniform potential, the gravitational force on the BEC can be canceled  using a magnetic field gradient \cite{3dbox}.
 It is of great interest to investigate the generation of vortex lattice in a rotating uniform BEC in a box potential and we undertake this challenge in this paper.  Specifically, we consider a square  or a circular box trap in the $x-y$ plane with a confining  wall at the boundary and a strong harmonic trap in the transverse $z$ direction, such that the dynamics in the $z$ direction can be integrated out \cite{luca} so as to yield a quasi-2D BEC in a box trap.  The confining wall at the boundary is taken to be very high, so that the condensate density outside the bounding box is effectively zero.

For a rapidly rotating quasi-2D BEC in a  box potential  composed of a fixed number of atoms with a fixed atomic scattering length, as expected, the number of generated vortices  increases with the increase of angular frequency of rotation $\Omega$.
For a square box potential,
the generated  vortices have a natural tendency to arrange in a square lattice, rather than a triangular lattice as in a rotating BEC in a harmonic trap \cite{fetter}.  A perfect square lattice of vortices  results  when the number of vortices is the square of integers, e.g. 1, 4, 9, 16, 25, 36, 49 etc., which can be arranged in a perfect square array.  This should be contrasted with the vortex lattice in a rotating  harmonically trapped BEC favoring a triangular lattice. When the number of vortices in the latter case is one of 1, 7, 19, 37, 61 etc. a perfect triangular lattice arranged in the form of a closed hexagonal shape results \cite{vors,vorl,theoryIT,theoryRT}.  Some deviation from a perfect square or triangular lattice results when the number of vortices lies between two sets of the above (magic) numbers.  

A  rotating  BEC in a circular box trap is  equivalent to a rotating super-fluid  He II in a cylindrical bucket as used in early experiments \cite{rs}. 
 In this case the number of generated vortices  increases linearly with $\Omega$ as suggested by Feynman \cite{feynman} for a rotating uniform super-fluid. The energy of the rotating BEC in a circular box is found to  decrease  with $\Omega$.
 For a circular box potential,    the generated vortices are found to lie on concentric closed orbits as was also the case for He II in a rotating bucket \cite{sf,cz}. For a small angular frequency of rotation,  the number of vortices is small occupying a single orbit. As the angular frequency of rotation is increased, more concentric orbits are needed to accommodate all the vortices.  Near the center these orbits have the shape of polygons, whereas  the outer orbits have 
circular shapes.  
  We demonstrate, employing real-time simulation under a small perturbation, 
   that the generated vortex lattices are dynamically stable.

In Sec. 2  the mean-field GP equation  for a   rapidly rotating  BEC, in a uniform box trap in the $x-y$ plane and 
a strong harmonic trap in the $z$ direction,
is presented.  Under a tight trap in the transverse direction, a quasi-2D version of the model is  derived \cite{luca}, which we use in the present study. 
The results of numerical calculation are shown in Sec. 3, where we consider the generation of vortex lattice in a rapidly rotating quasi-2D uniform BEC in a square or circular box trap by solving the GP equation numerically employing the Crank-Nicolson  discretization scheme \cite{imag}.  
Finally, in Sec. 4 we present a brief summary of our findings.

\section{Mean-field model for a rapidly rotating  uniform  quasi-2D BEC}

As in a rotating super-fluid, the integral (\ref{qv}) over 
the closed path ${\cal C}$ is nonzero, it implies  topological defect inside this path, and the domain where $\bf v$ is
well defined is multiply connected. In the problem of a rotating super-fluid,  the  topological defect
is the quantized vortex.  The quantization of circulation can be
explained  assuming that the dynamics of super-fluids is driven by a complex scalar field \cite{fetter,phase,ngp}
\begin{equation}
\phi({\bf r},t)= |\phi({\bf r},t)|   e^{i \theta({\bf r},t) },
\end{equation}
 which satisfies the nonlinear mean-field GP equation \cite{fetter,ngp}
\begin{eqnarray}\label{gpeq}
{\mbox i} \hbar \frac{\partial \phi({\bf r},t)}{\partial t}  &=&
{\Big [}  -\frac{\hbar^2}{2m}\nabla^2
+ \frac{1}{2}m \omega^2 
{z}^2+V({\bf r} ) \nonumber \\ &+&  \frac{4\pi \hbar^2}{m}{a} N \vert \phi({\bf r},t)\vert^2
{\Big ]}  \phi({\bf r},t),
\end{eqnarray}
where   $m$ is the mass of an atom, 
$N$ is the number of atoms, $V({\bf r})$ is the  trapping potential, and   $a$ is the atomic scattering length. The function $\phi$ is normalized as $\int d{\bf r}|\phi({\bf r},t)|^2 =1.$

A rapidly rotating BEC is conveniently described  in the rotating frame, as in that  frame of reference  the generated vortex  
lattice is a stationary state \cite{fetter}. The  vortex lattice  can be obtained numerically  by
solving the underlying mean-field GP equation of the trapped BEC in the rotating frame by the imaginary-time propagation method \cite{imag}. We consider a uniformly trapped BEC in the $x-y$ plane with a strong 
harmonic trap of angular trapping frequency $\omega$ in the $z$ direction. 
 To write the 
dynamical equation of the trapped BEC in the rotating frame,  we note that the Hamiltonian in the rotating frame is given by \cite{ll1960} $H = H_0-\Omega l_z$, where $H_0$ is the same  in the laboratory frame, with $\Omega $   the angular frequency of rotation,  $l_z$  the $z$ component of angular momentum given by $l_z \equiv {\mbox i}\hbar (y\partial/\partial  x - x \partial/\partial y )$.
The above transformation to the rotating frame suggests that the energy of the ground state
of a uniform BEC  in the rotating frame should decrease  as the angular frequency of rotation is increased  as will be verified in our numerical calculations \cite{feynman,fetter}.     
With the inclusion of the extra rotational energy  $-\Omega l_z$ in the Hamiltonian,   the mean-field 
GP equation (\ref{gpeq}) for the trapped BEC in the rotating frame for $\Omega <\omega $  can be written as \cite{fetter}
\begin{eqnarray} \label{eq1x} 
{\mbox i} \hbar \frac{\partial \phi({\bf r},t)}{\partial t}  &=& 
{\Big [}  -\frac{\hbar^2}{2m}\nabla^2 -\Omega l_z 
+ \frac{1}{2}m \omega^2 
{z}^2+V(x,y) \nonumber \\ &+&  \frac{4\pi \hbar^2}{m}{a} N \vert \phi({\bf r},t)\vert^2
{\Big ]}  \phi({\bf r},t),
\end{eqnarray}
 where $\frac{1}{2}m \omega^2 {z}^2$ is the harmonic trapping potential along 
$z$ direction.
  We consider two forms of the box potential   $V(x,y)$ in  (\ref{eq1x}):  a square box and a circular box. 
The square box potential is given by 
\begin{eqnarray}\label{v1}
 V(x,y) &= & 0,  \quad -d/2<x,y<d/2\\
 V(x,y) & \to & \infty, \quad  -d/2 >x,y> d/2 \label{v2}
\end{eqnarray}
where $x=\pm d/2 $ and 
 $y=\pm  d/2
$  define the boundaries of the box potential.    The circular box potential is taken as  
   \begin{eqnarray}\label{v3}
 V(x,y) &= & 0,  \quad \sqrt{x^2+y^2}<{\cal R},\\
 V(x,y) & = & V_0, \quad  \sqrt{x^2+y^2}\ge {\cal R}, \label{v4} 
\end{eqnarray} 
where $\cal R$ is the radius of the circular box potential. In this case, for numerical convenience, we consider a large finite value of $V_0$ $ (=60)$  for $  \sqrt{x^2+y^2}\ge {\cal R}$. As most numerical simulations are performed on a spatial  square mat in Cartesian space, the potential has to be finite everywhere. 
{ This is why, in the case of a circular box potential, we take a finite large value of $V_0$ ($=60$) outside the box.  Provided that $V_0$ is large compared to the energy of the system, the result will be independent of the value of $V_0$ employed. 
We verified that a larger value of $V_0$ does not change the generated vortex lattice for a circular box potential.}
For the square box potential, the  square mat is taken identical to the size of the box potential and the wave function is taken to be zero on the boundary. For the circular box potential of radius $\cal R$, a square mat of size $2{\cal R}$ is considered. 
However, this approach of describing the rotating BEC using the mean-field GP equation (\ref{eq1x}) in the rotating frame 
has a limited validity. If the angular  frequency $\Omega$ is increased beyond the trapping frequency $\omega$, the rotating BEC makes a quantum phase transition to a non-super-fluid state, where a mean-field description of the rotating BEC might not be valid \cite{fetter}.

The following   dimensionless form of  (\ref{eq1x})  can be obtained  by  the  transformation of variables: ${\bf r}' = {\bf r}/\hat l, \hat l\equiv \sqrt{\hbar/m\omega}$, $t'=t\omega,  \phi_i'=   \phi_i \hat l^{3/2},  \Omega'=\Omega/\omega, l_z '= l_z/\hbar$ etc.:   
\begin{eqnarray}& \,
{\mbox i} \frac{\partial \phi({\bf r},t)}{\partial t}=
{\Big [}  -\frac{\nabla^2}{2 }
+V(x,y)+ \frac{1 }{2}  z^2  -\Omega  l_z 
+ 4\pi Na \vert \phi \vert^2 
{\Big ]}  \phi({\bf r},t),
\label{eq3} 
\end{eqnarray}  
where for simplicity we have dropped the prime from the transformed variables.

For a quasi-2D binary BEC in the $x-y$ plane, under a strong trap along the $z$ 
direction, the essential vortex dynamics will be confined to  the $x-y$ plane with the $z$ dependence playing a passive role.   The wave functions  can then be written as 
$\phi({\bf r},t)= \psi({x,y};t)\Phi(z)$, where the function $ \psi({x,y};t)$ carries the essential vortex dynamics and $\Phi(z)$ is the normalizable Gaussian function   $\Phi(z)= exp(-z^2/2d_z^2)/(\pi d_z^2)^{1/4}$. In this case the 
$z$ dependence can be integrated out \cite{luca} and we have the following 2D equations
\begin{eqnarray}
{\mbox i} \frac{\partial \psi({x,y};t)}{\partial t} &=&
{\Biggr [}  -\frac{\nabla^2}{2 }
+V(x,y) -{\mbox i}\Omega \Big (y \frac{\partial}{\partial x}-  x \frac{\partial}{\partial y}  \Big )\nonumber \\
&+& g \vert \psi \vert^2  
{\Biggr ]}  \psi({x,y};t),
\label{eq5} 
\end{eqnarray}
where
$g=2\sqrt{2\pi} a N/d_z$, and normalization $\int | \psi({x,y})|^2 dx dy =1$.    In this study we will   consider $\Omega<1$ \cite{fetter}. This reduction to a quasi-2D form of the GP equation is also possible in the case of  a uniform density along the $z$ direction, e.g. $\Phi(z)$ is a constant between two finite $z$ values ($-z_0<z<z_0$), in place of a strong trap.  In that case nonlinearity $g$ will be $g=4\pi N a \int |\Phi(z)|^4 dz / \int |\Phi(z)|^2 dz$.

To evaluate the energy of the BEC in the rotating frame, we note that the wave function $\psi(x,y;t)$
is intrinsically complex. Hence in place of  
evaluating the real energy from ~(\ref{eq5})   
involving complex algebra over complex wave function, it is convenient to write a real expression for the energy. To calculate the energy, we write the two coupled non-linear equations for the real and imaginary parts of the wave function $\psi=\psi_R+\mathrm i \psi_I\equiv \sqrt{\psi_R^2+\psi_I^2}\exp(i\varphi)$, where $\varphi$ is the phase of the wave function, viz.~ (2.1) of Ref.~\cite{jeng}. The equation satisfied by the real part is 
 \begin{eqnarray}\label{real}
\mathrm i\frac{\partial \psi_R({x,y};t)}{\partial t} &=& \left[-\frac{1}{2}\nabla^2+V(x,y) +g|\psi({x,y};t)|^2\right] \psi_R({x,y};t) \nonumber \\
&+&\Omega\left( y\frac{\partial} {\partial x} -x \frac{\partial}{\partial y} \right) \psi_I({x,y};t)\, .
\end{eqnarray}
In this equation $\psi_R$ is not normalized to unity.
Using ~(\ref{real}), the energy per atom in the rotating frame  can be expressed  as 
\begin{eqnarray}\label{e2dr}
E&=& \frac{1}{\int dx dy  \psi_R^2 } \int dx dy   \biggr[-{\frac{1}{2}{(\nabla \psi_R)^2}}+ V(x,y)\psi_R^2+
 \frac{1}{2} g(\psi_R^2+\psi_I^2){\psi_R^2}  \nonumber \\ &+&\Omega    \psi_R \left( y\frac{\partial }{\partial x} -
x \frac{\partial}{\partial y} \right)\psi_I\biggr] .
\end{eqnarray}
Equation~(\ref{e2dr})  involves algebra of real functions only. Hence this approach leads to far more accurate numerical result.

\section{Numerical Results}

\begin{figure}[!t]

\begin{center}
\includegraphics[trim = 0cm 0.cm 0cm 0mm, clip,width=\linewidth]{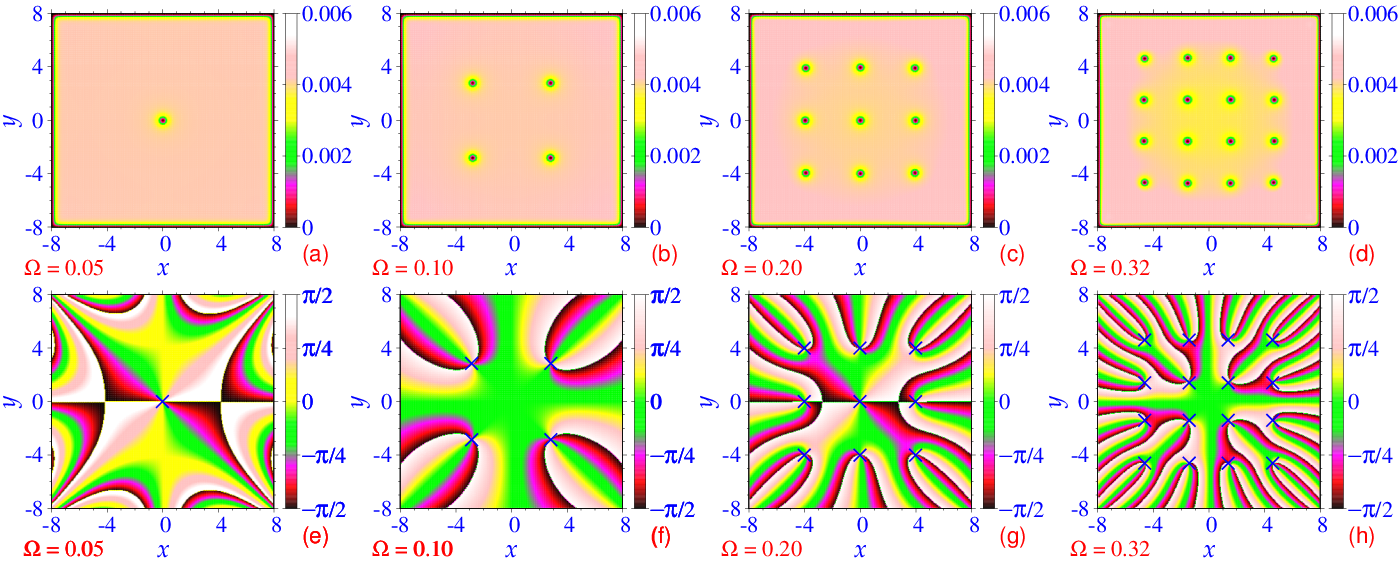} 
 
\caption{  Vortex lattices in a rapidly rotating quasi-2D BEC with $g=5000$, satisfying  (\ref{eq5}),   in a square box, viz.  (\ref{v1}) and (\ref{v2}), 
 of size $d=16$  from a contour plot of 2D densities ($|\psi|^2$) for $\Omega =$  (a) $0.05$, (b) 0.10, (c) 0.20, and (d) 0.32. The numbers of vortices in these plots are squares of integers. 
 (e)-(h) display the  corresponding  phase profiles $\varphi \equiv \arctan (\psi_I/\psi_R)$ of the rotating BECs illustrated in (a)-(d),  respectively.  The crosses in (e)-(h) show the position of the vortices in the rotating BEC. All quantities plotted in this and following figures are dimensionless.
}
\label{fig1}
\end{center}

\end{figure}

The  mean-field equation for a quasi-2D BEC in a rotating box trap 
 (\ref{eq5})  cannot be solved analytically and different numerical methods, such as the split time-step Crank-Nicolson method \cite{imag,CPC} or the pseudo-spectral method \cite{PS}, are usually  employed  for
its solution. Here we solve   (\ref{eq5})  by the split time-step
Crank-Nicolson discretization scheme using a space step of 0.05
and a time step of 0.00025.  There are different
C and FORTRAN programs for solving the GP equation \cite{imag,CPC}  and one should use the appropriate one.
These programs have recently been adapted to simulate the vortex lattice in a rapidly rotating BEC \cite{cpckk} and we use these in this study. 
 In this paper, without considering a specific atom, we will present the results in dimensionless units for different sets of  parameters: 
$\Omega, g  $.  As the mean-field  equation  (\ref{eq5}) refers to the rotating frame, where the BEC wave function is a stationary state, we can use the imaginary-time propagation method to generate the minimum-energy vortex lattice.   The imaginary-time propagation was started with the following Gaussian and/or  one-vortex initial states
\begin{equation}\label{sample}
\Psi(x,y)= \frac{1}{\sqrt{\pi w^2}}\exp\biggr [ -\frac{x^2+y^2}{2w^2}  \biggr ],  \quad
\Psi(x,y)= \frac{x+\mbox i y}{\sqrt{\pi w^2}}\exp\biggr [ -\frac{x^2+y^2}{2w^2}  \biggr ],
\end{equation}
where $w$ is the width. { It is well known \cite{kkk} that the generated vortex lattice in imaginary-time simulation using a phase-correlated initial state, such as (\ref{sample}), is sensitive to the particular initial state used.  Usually, there are many  meta-stable states near the true ground state and the numerical simulation as well as  the experiments may settle to one of these meta-stable states. To minimize this problem to a great extent, following the suggestion in Ref. \cite{cpckk}, we multiply the initial states (\ref{sample}) by a random phase at each space grid point and use the  functions $  \Psi_R(x,y)$ with a random phase  as the initial states in actual calculation:
\begin{equation}\label{random}
\Psi_R(x,y)= \Psi(x,y)\exp\big[- 2 \pi {\mbox i} { R}(x,y)  \big],
\end{equation}  
where $ { R}(x,y)$ is a random number between 0 and 1. If the initial state (\ref{random}) is used in numerical simulation, any initial integrable function, viz. (\ref{sample}), leads to the true ground state when the number of vortices is small. But when the number of vortices increases to a very large number, there will be many meta-stable states with energy close to that of the ground state and  there is a probability that the imaginary-time simulation will converge to one of these meta-stable states instead of the true ground state. 
In this paper we present the vortex lattice states with the minimum  of energy.   }
     
\begin{figure}[!t]

\begin{center}
\includegraphics[trim = 0cm 0cm 0cm 0mm, clip,width=.8\linewidth]{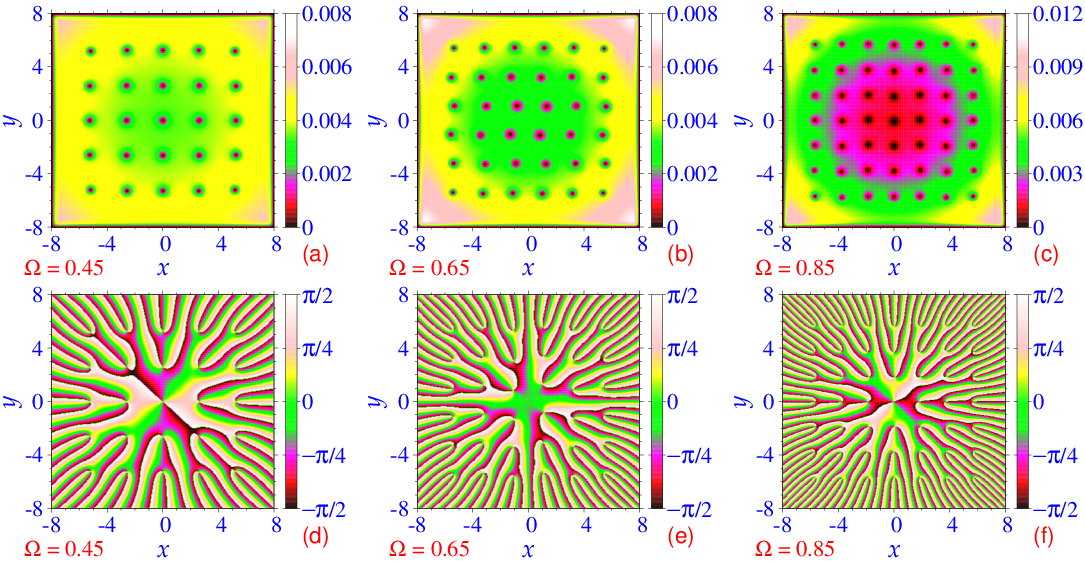} 
 
\caption{    Vortex lattices in a rapidly rotating quasi-2D BEC with non-linearity $g=5000$, satisfying  (\ref{eq5}),  in a square box of size $d=16$  from a contour plot of 2D densities ($|\psi|^2$) for $\Omega =$  (a) $0.45$, (b) 0.65, and (c) 0.85.  The numbers of vortices in these plots are squares of integers.  (d)-(f) display the  corresponding  phase profiles $\varphi \equiv \arctan (\psi_I/\psi_R)$ of the rotating BECs illustrated in (a)-(c),  respectively.
}
\label{fig2}
\end{center}

\end{figure}

First we consider the generation of square vortex lattice in a rapidly rotating quasi-2D BEC confined by a square box potential of size $d =16$ for a fixed non-linearity $g=5000$ in  (\ref{eq5}), viz.  (\ref{v1}) and (\ref{v2}). As the angular frequency of rotation $\Omega$ is increased, the BEC generates more and more vortices.  We find that, when the total number of vortices is the square of an integer, the vortices have a natural tendency to arrange in a square lattice.   This is illustrated in figures \ref{fig1}(a)-(d) and   \ref{fig2}(a)-(c) for $\Omega= 0.05, 0.10, 0.20, 0.32, 0.45, 0.65$, and 0.85. The corresponding number of vortices are $1, 4,9,16,25,36,$ and 49, respectively, which are the squares of integers $1, 2, 3, 4, 5, 6, 7$.  The individual vortices in these plots are of unit angular momentum, which can be verified from a consideration of phase    $\varphi=\arctan (\psi_I/\psi_R)$  of the wave function $\psi$ in the BEC.  A complete rotation in a close contour around a vortex of unit angular momentum should generate a phase of $2\pi$ according to $\psi \sim e^{i\varphi}.$ The corresponding phase profiles of the rotating BEC are illustrated in figures \ref{fig1}(e)-(h).  From the plots of phase profiles in figures \ref{fig1}(e)-(h), it is verified that each vortex carry an unit of angular momentum. Similar phase profiles of the vortices in figures \ref{fig2}(a)-(c) are shown in figures \ref{fig2}(d)-(f).

\begin{figure}[!t]

\begin{center}
 
 \includegraphics[trim = 0cm 0cm 0cm 0cm, clip,width=\linewidth]{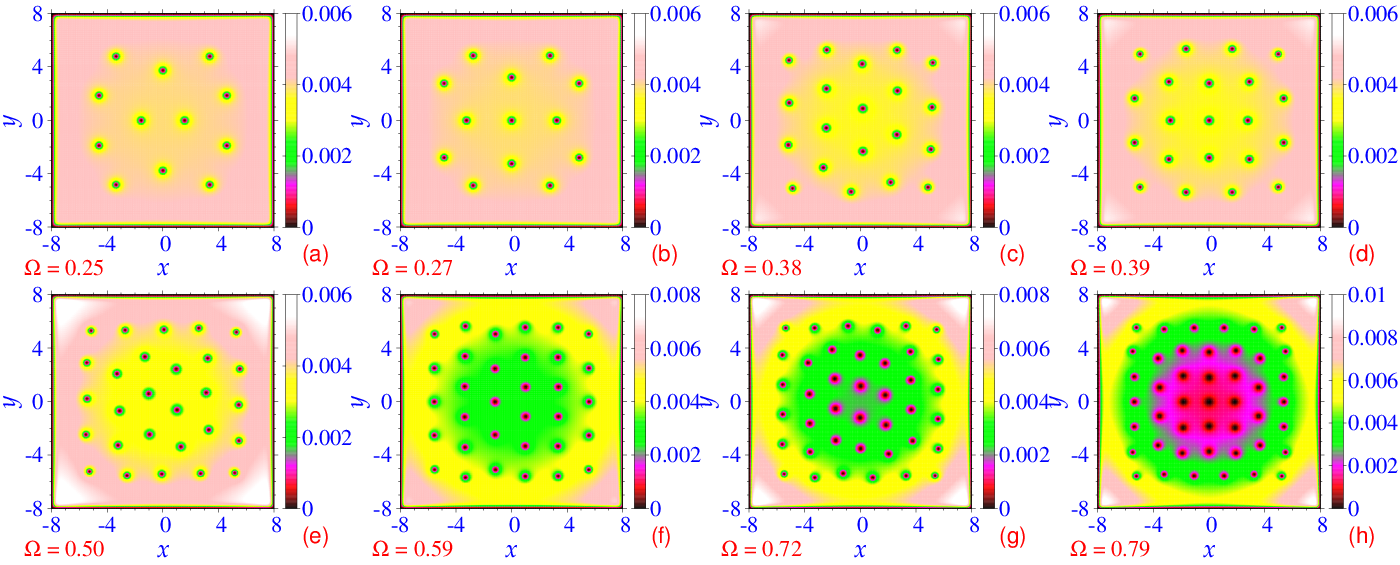} 
\caption{  Vortex lattices in a rapidly rotating quasi-2D BEC with $g=5000$  in a square box of size $d16$  from a contour plot of 2D densities ($|\psi|^2$) for $\Omega =$  (a) $0.25$, (b) 0.27, (c) 0.38,  (d) 0.39, (e) 0.50, (f) 0.59, (g) 0.72, and (h) 0.79. The number of vortices in these plots are 12, 13, 20, 21, 28,  33, 40, and 45, which, different from figures \ref{fig1} and \ref{fig2}, are not squares of integers.
}
\label{fig4}
\end{center}

\end{figure}

The angular frequency of rotation $\Omega$ in figures \ref{fig1} and \ref{fig2}  were chosen appropriately  to generate vortex lattices, where the number of vortices are squares of integers and the vortices are arranged on a perfect square lattice. However, for an  arbitrary  $\Omega$, the number of generated vortices are not squares of integers.   Consequently, in these cases no definite perfect lattice was found in general. This is illustrated in figure \ref{fig4}  where we exhibit the vortex lattices for $g=5000$ and $\Omega =$  0.25, 0.27, 0.38,  0.39, 0.50,  0.59, 0.72, and 0.79. For small $\Omega$ in figures \ref{fig4}(a)-(c) no definite lattice structure is found.  In figures   \ref{fig4}(d) and (h) an approximate square lattice can be seen.  However, in figures  \ref{fig4}(e), (f) and (g) an approximate hexagonal lattice can be identified near the center inside a square boundary. 

 The number of generated vortices  in the rotating BEC  can be calculated using a theoretical estimate due to Feynman \cite{feynman}. A rapidly rotating uniform condensate has a dense array of vortices,
with a uniform areal density     (number of vortices per unit area)   \cite{fetter}
 \begin{equation}
{\cal N}= \frac{\Omega}{\pi},
\end{equation}
in units $m=\hbar =1$. In these units, in a circular uniform quasi-2D BEC,  the space available for a single vortex is  $\pi /\Omega $.   
{ This sets an estimate for the distance between cores of  vortices as $2/\sqrt \Omega$ and this estimate is approximately valid for all vortex lattices illustrated in this paper. }
  Feynman's estimate   for the total number of vortices in a BEC of area $\cal A$  is
${\cal N} = {\cal A} \Omega /\pi $. Hence the total number of vortices  for a square of side $d$ and for a circle of radius ${\cal R}$   are  
 \begin{eqnarray}\label{feynmans}
{\cal N}_{\mbox sq} &=&  \frac{d^2  \Omega}{\pi},   \\
  \label{feynmanc}
{\cal N}_{\mbox ci} & =&  {\cal R}^2  \Omega,
 \end{eqnarray}
respectively. In this paper we use $d=2{\cal R}.$
{The Feynman estimates (\ref{feynmans})  and (\ref{feynmanc}) show that the number
of vortices   increases linearly with $\Omega$ and give an idea about  how many vortices are going to be generated in an actual numerical simulation.}

\begin{figure}[!t]

\begin{center}
\includegraphics[trim = 0cm 0.0cm 0cm 0mm, clip,width=.9\linewidth]{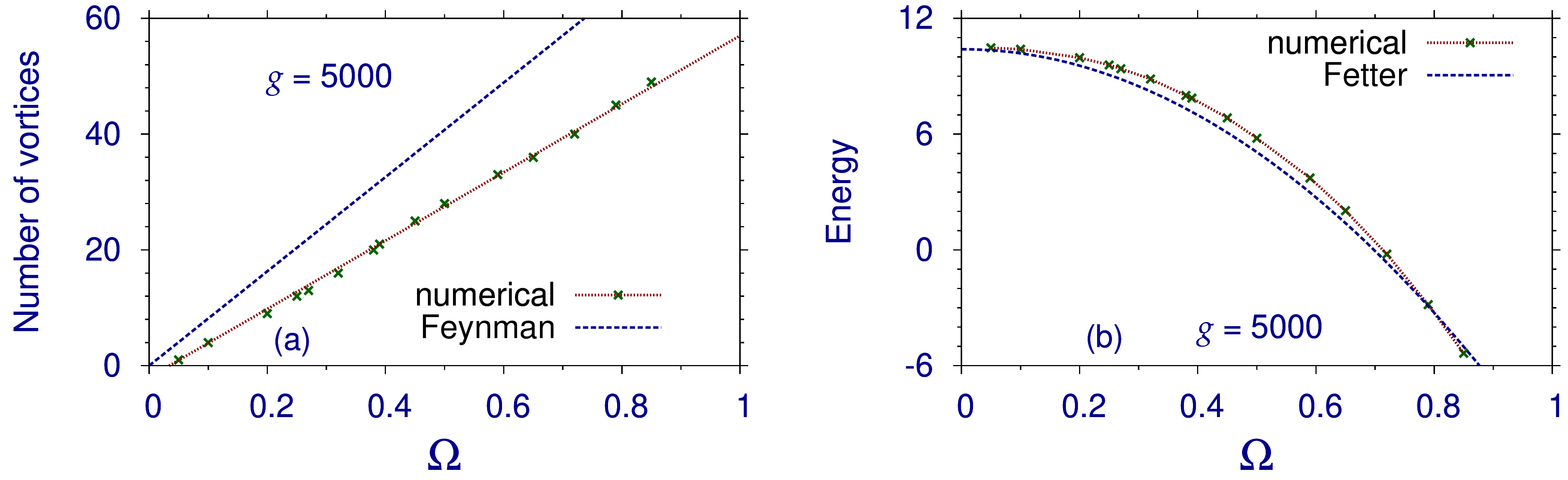} 
\caption{ (a)  Number of vortices and (b) energy per atom  in the rotating frame for a rapidly rotating quasi-2D BEC 
confined in a square box  with $g=5000$ versus angular frequency of rotation $\Omega$.  The theoretical estimates for number and energy (\ref{feynmans}) and (\ref{feynmanse}) with $E_{0\mbox{sq}}=10.4$
due to Feynman and Fetter are also shown.  
 The crosses are the actual points obtained numerically whereas the full lines are shown to guide the eye. In these displays, in addition to vortex lattices of figures \ref{fig1} and \ref{fig2}, where the numbers of vortices are the squares of integers, those of figure \ref{fig4}, with an arbitrary number of vortices,  are also plotted.
}
\label{fig3}
\end{center}

\end{figure}

An estimate of the $\Omega$-dependent part of the energy in the rotating frame can be obtained following Fetter 
  \cite{fetter}.  The relevant part of the classical energy density in the rotating frame, per atom of mass $m=1$, moving with velocity ${\bf v(r)}$,  is  $(v^2/2-{\bf \Omega} \cdot {\bf r\times v})\rho$, where $\rho$ is the density. For rotation with angular frequency $\Omega$,  ${\bf v=\Omega  \times r}$. Using this result 
and recalling that ${\bf \Omega \cdot  r\times   (\Omega \times r}) =|{\bf \Omega \times r} |^2$, we get for 
the  $\Omega$-dependent part of the energy per atom as $-\int d {\bf r} |{\bf \Omega \times r}|^2 \rho/2= - I \Omega^2/2$, where $I$ is the equivalent moment of inertia of  solid-body rotation of the superfluid. The total energy should then have the form $E=E_0- I \Omega^2/2 $, where $E_0$ is the  energy of the non-rotating BEC with $\Omega=0$  
 in Eq. (\ref{e2dr}). For a quasi-2D BEC atom  of mass $m=1$, the moment of inertia of a square of side $d$ is
$ d^2/6= 2{\cal R}^2/3$, as $d=2{\cal R}$,   and that of a circle of radius $\cal R$ is  ${\cal R}^2/2$. 
Hence the Fetter estimates of energies {\it per atom}  of a quasi-2D square-shaped and circular BECs are, respectively,
\begin{eqnarray}\label{feynmanse}
{E}_{\mbox{sq}} &=& E_{0\mbox{sq}}- \frac{{\cal R}^2\Omega^2}{3 },   \\
  \label{feynmance}
{E}_{\mbox{ci}} & =& E_{0\mbox{ci}} - \frac{{\cal R}^2\Omega^2}{4 },
 \end{eqnarray}
where $E_{0\mbox{sq}}$ and $ E_{0\mbox{ci}} $ are the respective energies for $\Omega=0$.
It is remarkable that the rotational energies in Eqs. (\ref{feynmanse}) and (\ref{feynmance}) are 
parameter free.

It is interesting to investigate how the number of vortices in the rapidly rotating quasi-2D BEC confined in a box trap increases in numerical simulation as the angular frequency of rotation $\Omega$ is increased. In figure \ref{fig3}(a) 
we display  the number of vortices as a function of the angular frequency of rotation. In this plot we also considered angular frequencies for which the number of vortices is not a square of an integer. Specifically, we include all the vortex lattices shown in figures \ref{fig1}, \ref{fig2} and \ref{fig4}.   The number of vortices increases with $\Omega$.  We also show in this plot the parameter-free  Feynman estimate (\ref{feynmans}) for the number of vortices.  
Nevertheless, the energy per atom in the rotating frame (\ref{e2dr}) decreases with the increase of angular frequency of rotation $\Omega$ as shown in figure \ref{fig3}(b) for the same vortex-lattice states exhibited in figure \ref{fig3}(a).  The Fetter estimate for energy   (\ref{feynmanse}) with $E_{0\mbox{sq}}=10.4$ is also shown in this plot.

\begin{figure}[!t]

\begin{center}
 
 \includegraphics[trim = 0cm 0.64cm 0cm 0cm, clip,width=\linewidth]{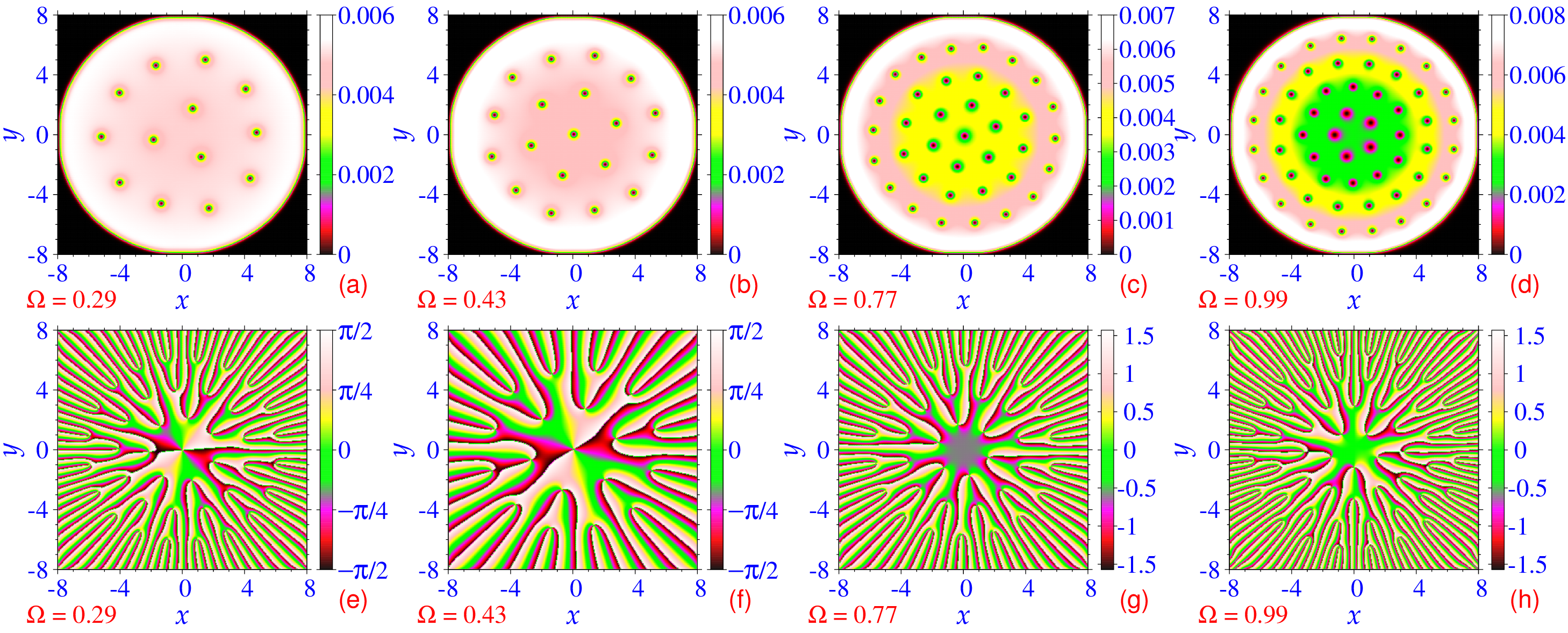} 
\caption{  Vortex lattices in a rapidly rotating quasi-2D BEC with $g=5000$  in a circular   box, viz.  (\ref{v3}) and (\ref{v4}), 
 of radius ${\cal R}=8$  from a contour plot of 2D densities ($|\psi|^2$) for $\Omega =$  (a) $0.29$, (b) 0.43, (c) 0.77, and (d) 0.99.  (e)-(h) display the  corresponding  phase profiles $\varphi  $ of the rotating BECs illustrated in (a)-(d),  respectively.  The number of generated vortices are (a) 13, (b) 19, (c) 37, and (d) 53.
}
\label{fig5}
\end{center}

\end{figure}

\begin{figure}[!t]

\begin{center}
 
 \includegraphics[trim = 0cm 0.64cm 0cm 0cm, clip,width=\linewidth]{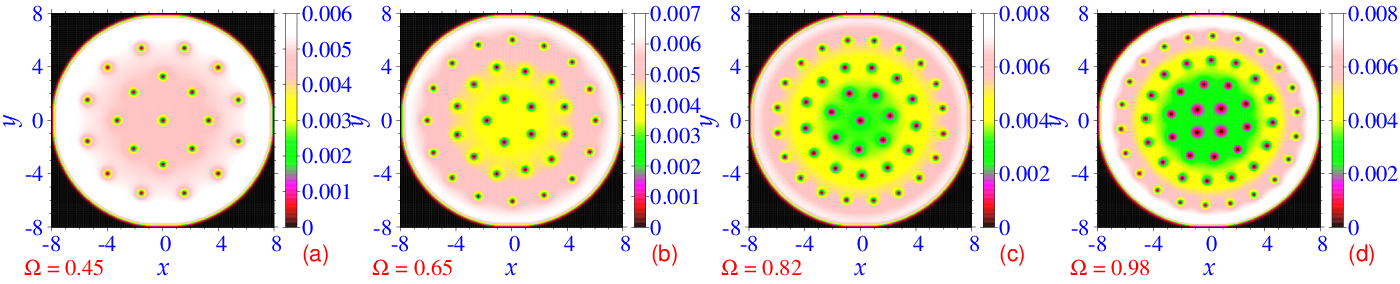} 
\caption{  Vortex lattices in a rapidly rotating quasi-2D BEC with $g=5000$  in a circular   box, viz.  (\ref{v3}) and (\ref{v4}), 
 of radius ${\cal R}=8$  from a contour plot of 2D densities ($|\psi|^2$) for $\Omega =$  (a) $0.45$, (b) 0.65, (c) 0.82, and (d) 0.98.   The number of generated vortices are (a) 21, (b) 33, (c) 41, and (d) 52.
}
\label{fig6}
\end{center}

\end{figure}

The formation of vortex lattice in a circular box potential of radius ${\cal R}=8$, viz.  (\ref{v3}) and (\ref{v4}), is considered next for non-linearity $g=5000$. The nonlinear equation (\ref{eq5}) was discretized  in a square box of size $d=16$ for numerical simulation.
 In this case the vortices naturally arrange in closed concentric orbits. The orbits have shape of polygons (square, pentagon, hexagon, heptagon, etc.) near the center of the BEC, which change to near circular shape near the periphery. 
  This is illustrated in figure \ref{fig5} where we display the vortex lattices for $\Omega=$ (a) 0.29, (b) 0.43, (c) 0.77, and (d) 0.99  and the corresponding phase distribution in plots (e), (f), (g), and (h), respectively. In figures \ref{fig5} (a), (b), (c), and (d) we find 1, 2, 3, and 4 concentric orbits  on which the vortices lie. In plots  (b) and (c) there is a vortex at the center, whereas in plots (a) and  (d) the central spot is vacant.   In figure \ref{fig6} we display the vortex lattices for $\Omega=$ (a) 0.45, (b) 0.65, (c) 0.82, and (d) 0.98, where in plots (a) and (c) there is a vortex at the center and in plots (b) and (d) the central spot is vacant. 
 {    
Although the Feynman estimate gives the total number of vortices, it cannot it cannot predict if the central spot will be occupied or not.  The position and the distribution of the vortices are determined by the condition of minimization of energy. }

In figures \ref{fig5}(b) and (c), the vortices are arranged on slightly deformed triangular Abrikosov lattice \cite{abri}. In both these figures a closed hexagon, reminiscent of triangular lattice, can be identified at the center, which is surrounded by another slightly rounded closed hexagon in figure \ref{fig5}(b), whereas in figure \ref{fig5}(c) the same is surrounded
by two closed rounded  hexagons, as in a triangular lattice. However, in figure \ref{fig5}(c) 
the outermost orbit has practically a circular shape. The underlying hexagonal structure  in these plots can be realized upon a closer look and counting the number of vortices on each orbit, e.g. 1, 6, 12, and 18.  
The other vortex lattices in figures \ref{fig5} and \ref{fig6} do not exhibit triangular-lattice structure.  Similar conclusion was also made fifty years ago in a circularly symmetric rotation of a  super-fluid in a bucket \cite{sf,cz}.   In a numerical study of rotating He II in a cylindrical container Stauffer and Fetter \cite{sf}  also found that the vortices lie in concentric circles around the center and  in some cases, some vortices lie on a triangular lattice near the center.

\begin{figure}[!t]

\begin{center}
 
 \includegraphics[trim = 0cm 0cm 0cm 0cm, clip,width=\linewidth]{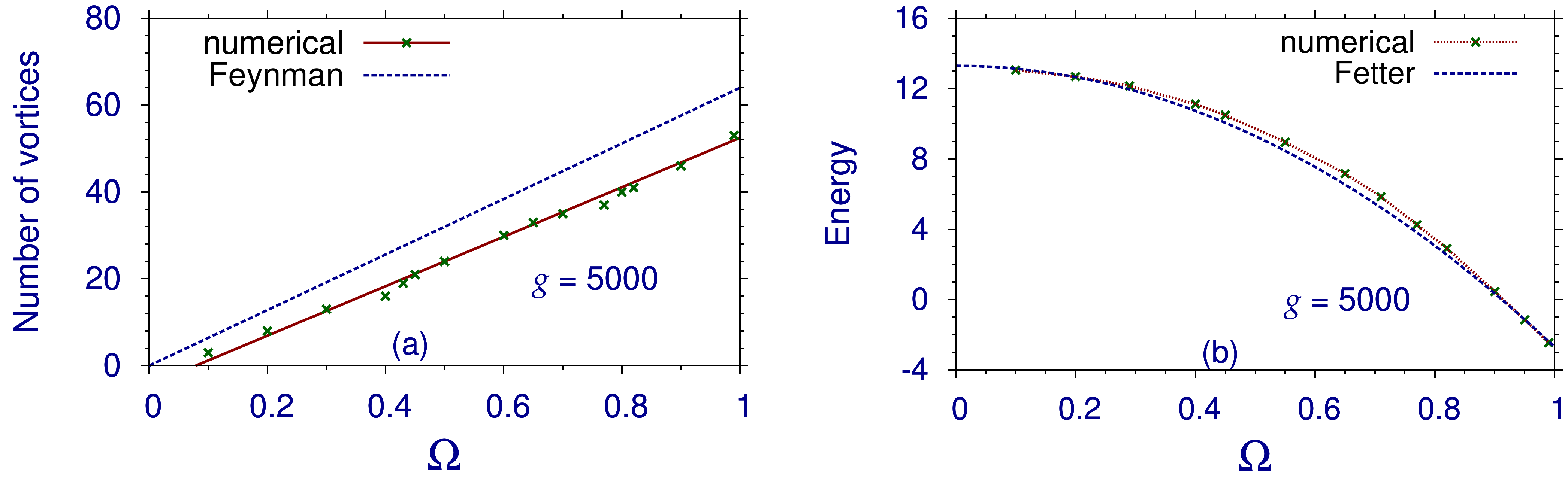} 
\caption{ (a)  Number of vortices and (b) energy per atom  in the rotating frame for a rapidly rotating quasi-2D BEC 
confined in a circular box of radius ${\cal R}=8$  with $g=5000$ versus angular frequency of rotation $\Omega$.
 The theoretical estimates for number and energy (\ref{feynmanc}) and (\ref{feynmance}) with $E_{0{\mbox{ci}}}=13.3$ due to Feynman and Fetter are also shown.  
The crosses are the actual points obtained numerically whereas the straight  lines are shown to guide the eye.  
  }
\label{fig7}
\end{center}

\end{figure}

The number of vortices and the respective energies obtained in numerical simulation of a BEC in a circular box trap is considered next. In figures \ref{fig7}(a) and (b) we plot the numerically obtained number of vortices and energy per atom versus angular frequency of rotation and compape these with the respective Feynman and Fetter estimates (\ref{feynmanc})  and   (\ref{feynmance})  with $E_{0{\mbox{ci}}}=13.3$. It is remarkable that the parameter-free Fetter estimate of $\Omega$-dependent  energy in the rotational frame agrees so well with actual numerical simulation.

{A balance between the kinetic energy ($\sim 1/2\xi^2$)
and the interaction energy ($\sim g |\psi(x,y)|^2$), viz. Eq. (\ref{eq5}),  of a BEC 
leads to a typical length scale called coherence length $\xi \equiv (2g |\psi(x,y)|^2)^{-1/2}$ 
for a weakly interacting BEC \cite{smith}. This quantity is relevant for
super-fluid effects. For instance, it provides the typical size of the core of quantized vortices \cite{abri,smith,gross}.
 Under rotation, like a classical centrifugal action, most atoms of the BEC move to the periphery thus increasing the density  near the boundary, viz. figures \ref{fig5} and \ref{fig6}.  The density is high near the rigid wall, where it abruptly falls to zero as can be seen in  figures \ref{fig5} and \ref{fig6} without any numerical consequence. 
 The coherence length being proportional to $ |\psi(x,y)|^{-1}$  is thus smaller near the periphery. As vortex core radius is determined by the coherence length \cite{smith}, the radius of the vortex core is smaller near the boundary. For example, in numerical simulation  $|\psi(x,y)|^2$ is typically 0.002 near center and 0.008 near periphery. For $g=5000$, $\xi$ is typically 0.2 near the center and 0.1 near the boundary. 
Hence in   figures \ref{fig5} and \ref{fig6}  vortex spots are larger, and hence darker, near the center than the boundary.   }

\begin{figure}[!t]

\begin{center}
 
 \includegraphics[trim = 0cm 0.cm 0cm 0cm, clip,width=\linewidth]{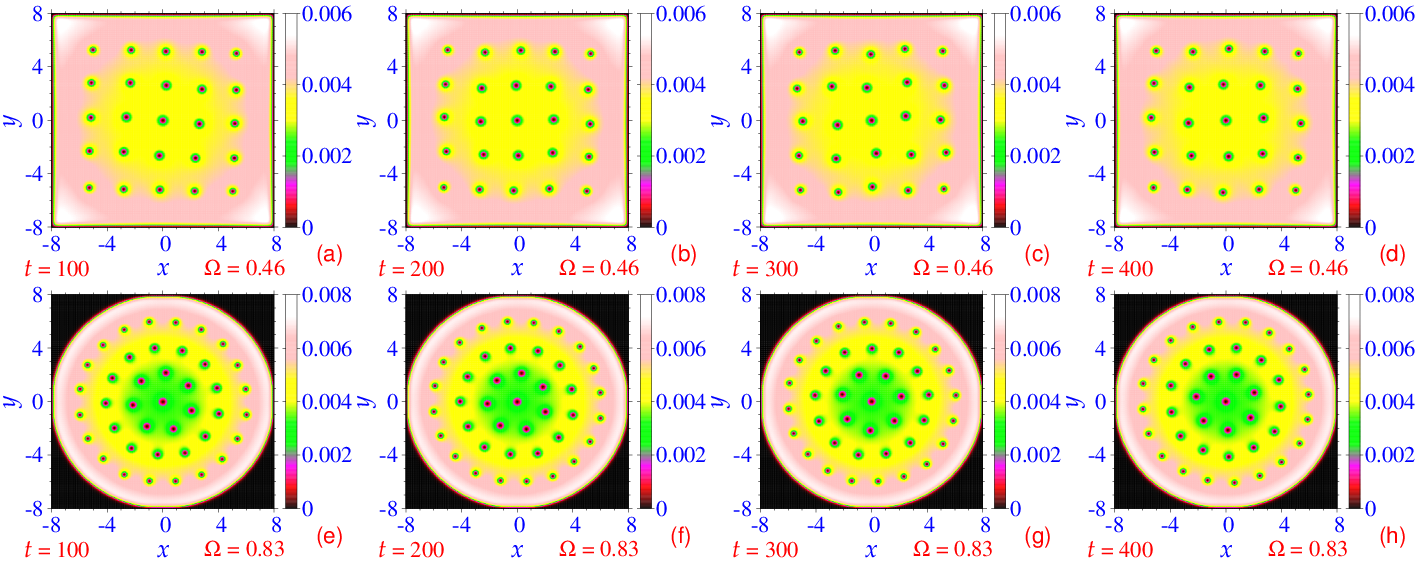}

\caption{  Dynamical evolution of  vortex lattice   of a rotating BEC in a square box, displayed in figure \ref{fig2}(a),
 during real-time propagation for 400 units of time using the corresponding imaginary-time wave function as input, at times (a) $t=100$,  (b) $t=200$, (c) $t=300$, and (d)  $t=400$. During real-time propagation the angular frequency of rotation $\Omega$ was changed at $t=0$  from the imaginary-time value of $\Omega = 0.45$ to 0.46. 
 Dynamical evolution of vortex lattice   of a rotating BEC in a circular box, shown in figure \ref{fig6}(c),
 during real-time propagation for 400 units of time using the corresponding imaginary-time wave function as input, at times (a) $t=100$,  (b) $t=200$, (c) $t=300$, and (d)  $t=400$. During real-time propagation the angular frequency of rotation $\Omega$ was changed at $t=0$  from the imaginary-time value of $\Omega = 0.82$ to 0.83. 
}
\label{fig8}
\end{center}

\end{figure}

The dynamical stability of the vortex lattices of the rotating uniform BEC in a square and circular box traps is tested next. For this purpose  we subject the vortex-lattice profiles of the rotating BEC to real-time evolution during a large interval of time,  after slightly changing the angular frequency of rotation $\Omega$ at $t=0$. The vortex lattice will be destroyed, if the underlying BEC wave function were dynamically unstable.  First we consider  real-time propagation of  the vortex lattice exhibited  in figure \ref{fig2}(a), after changing $\Omega$ from 0.45 to 0.46 at $t=0$. The consequent evolution  of the vortex lattice is displayed in  figure \ref{fig8} (a)  at $t=100$,           (b) $t=200$, (c) $t=300$, and (d)  $t=400$.
The  real-time propagation of  the vortex lattice exhibited  in figure \ref{fig6}(c), after changing $\Omega$ from 0.82 to 0.83 at $t=0$ is considered next.
 The consequent evolution  of the vortex lattice is displayed in  figure \ref{fig8} at (e) $t=100$,           (f) $t=200$, (g) $t=300$, and (h)  $t=400$. The robust nature of the snapshots of vortex lattice during real-time evolution upon a small perturbation, as exhibited in figure \ref{fig8}, demonstrates the dynamical stability of the vortex lattice.

\section{Summary and Discussion} 
 
We have studied the generation of    vortex lattices in a rapidly rotating quasi-2D uniform BEC  confined in a square or a circular box, such that the condensate density on the boundary is zero. In this study we solved the mean-field GP equation numerically using imaginary-time propagation  with Crank-Nicolson discretization. This set-up allows the study of super-fluidity through the  generation of vortex lattices under rapid rotation of a dilute ultra-cold uniform  BEC in a new environment not considered before.   Previous considerations of vortex-lattice generation were  limited to a harmonically trapped BEC.
 In the case of a circular box potential, the vortices appear on concentric orbits.  
Near the center, these orbits accommodate  a small number of vortices and have the shape of polygons. Near the periphery, the orbits  accommodate  a large number of vortices and have a nearly circular shape.
   In case of the square box potential, the generated vortices lie on a perfect square lattice when the number of vortices is the square of an integer, e.g., 1, 4, 9, 16, 25, etc.  When the number of vortices is arbitrary the vortices often lie on approximate square lattices. The number of vortices increases linearly with $\Omega $  and energy decreases quadratically  with $\Omega$  in both cases.  

 The present study of vortex-lattice generation  in a rapidly rotating uniform BEC  bounded by  a circular or a square-shaped   box 
allows us to test the Feynman's estimates  (\ref{feynmans}) and  (\ref{feynmanc}).  In agreement with Feynman's estimates,  the number of  vortices increase linearly with  $\Omega$ and  energy decreases linearly with  $\Omega$. The Fetter estimates for the $\Omega$-dependent part of energy (\ref{feynmanse}) and  (\ref{feynmance})
is found to be in excellent agreement with actual numerical simulation in  both cases.

We also demonstrated the dynamical stability of the generated vortex lattice, by steady  real-time simulation over a long period of time,  after introducing 
 a small perturbation by slightly changing $\Omega$.  
With present experimental know-how the present vortex lattices in a quasi-2D uniform BEC,   confined in a square or a circular box, can be tested in a laboratory.

\section*{Acknowledgements}
\noindent
SKA thanks Prof. A. L. Fetter for many interesting, pertinent and very helpful  comments on this investigation. 
SKA thanks the Funda\c c\~ao de Amparo \`a Pesquisa do
Estado de S\~ao Paulo (Brazil) (Project: 
2016/01343-7) and the Conselho Nacional de Desenvolvimento Cient\'ifico e Tecnol\'ogico (Brazil) (Project:
303280/2014-0) for partial support.

\end{document}